\documentclass[fleqn,10pt]{wlscirep}

\title{3-D phononic crystals with ultra-wide band gaps}

\author[1,*]{Yan Lu}
\author[2]{Yang Yang}
\author[2]{James K. Guest}
\author[1]{Ankit Srivastava}
\affil[1]{Department of Mechanical, Materials, and Aerospace Engineering, Illinois Institute of Technology, Chicago, IL, 60616,USA}
\affil[2]{Department of Civil Engineering, Johns Hopkins University, Baltimore, MD, 21218,USA}

\affil[*]{Corresponding author: ylu50@hawk.iit.edu}

\begin{abstract}
In this paper gradient based topology optimization (TO) is used to discover 3-D phononic structures that exhibit ultra-wide normalized all-angle all-mode band gaps. The challenging computational task of repeated 3-D phononic band-structure evaluations is accomplished by a combination of a fast mixed variational eigenvalue solver and distributed Graphic Processing Unit (GPU) parallel computations. The TO algorithm utilizes the material distribution-based approach and a gradient-based optimizer. The design sensitivity for the mixed variational eigenvalue problem is derived using the adjoint method and is implemented through highly efficient vectorization techniques. We present optimized results for two-material simple cubic (SC), body centered cubic (BCC), and face centered cubic (FCC) crystal structures and show that in each of these cases different initial designs converge to single inclusion network topologies within their corresponding primitive cells. The optimized results show that large phononic stop bands for bulk wave propagation can be achieved at lower than close packed spherical configurations leading to lighter unit cells. For tungsten carbide - epoxy crystals we identify all angle all mode normalized stop bands exceeding 100\%, which is larger than what is possible with only spherical inclusions.
 
\end{abstract}
\begin{document}

\flushbottom
\maketitle

\thispagestyle{empty}

\section*{Introduction}
There has been a recent surge of research effort towards achieving exotic dynamic behavior through novel microstructural design of periodic composites. Within mechanics and elastodynamics these responses can be categorized in two broad areas: phononics and metamaterials\cite{srivastava2015elastic}. Phononics is the study of stress wave propagation in periodic elastic composites. The phononic band-structure \cite{martinezsala1995sound} results from the periodic modulation of stress waves, and as such has deep similarities with areas like electronic band theory \cite{bloch1928quantum} and photonics \cite{ho1990existence}. These periodic modulations provide for very rich wave-physics and for the potential novel applications such as wave guiding\cite{khelif2003trapping}, ultrasound tunneling\cite{yang2002ultrasound}, acoustic rectification\cite{li2011tunable}, sound focusing\cite{yang2004focusing}, thermal property tuning \cite{zen2014engineering}, and novel wave refraction applications \cite{nemat2015refraction,nemat2015anti,srivastava2016metamaterial} (See \cite{hussein2014dynamics} for a comprehensive review). The definitive characteristic of a phononic crystal which distinguishes it from a homogeneous or randomly heterogeneous media is the existence of a frequency region where wave propagation is prohibited. This region, called the phononic band gap, directly or indirectly affects most of the proposed applications of phononic crystals. Therefore, it is of significant interest and impact to find out those phononic topologies for which the phononic band gap is very large. This is a tough computational problem, especially in 3-D, which requires the use of fast phononic solvers coupled with sophisticated topology optimization routines.

Topology optimization has evolved rapidly in recent years as a form-finding methodology for structural and materials design\cite{deaton2014survey,sigmund2013topology,cadman2013design,guest2016topology}. It seeks to optimize the distribution of material resources across a design domain such that a defined objective function is minimized (or maximized) and constraints satisfied. Typically, finite element methods are used to discretize the design domain and a material relative density $\rho_e$ ranging continuously from $0$ to $1$, is assigned to each element, with  $\rho_e=0$ and $\rho_e=1$ indicating the presence of only material 1 or material 2 in the element, respectively. Intermediate values represent mixtures of the two material phases and are prevented by penalizing their existence, such as through the Solid Isotropic Penalization Method\cite{bendsoe1989optimal, rozvany1991coc}.

Among the growing range of applications of topology optimization\cite{asadpoure2015topology,challis2014high}, there have been some recent applications of topology optimization on band gap structures. For photonic crystals Cox and Dobson\cite{dobson1999maximizing, cox2000band} applied topology optimization to maximize band gaps in two-dimensional photonic crystals for $E$ and $H$ polarization. Jensen and Sigmund\cite{jensen2004systematic} presented results for optimized 2-D photonic waveguide design. Rupp et al.\cite{rupp2007design} presented optimization of 3-D surface wave guide. Robust topology optimization considering manufacturing variations in 2D photonic crystals were proposed by Wang et al.\cite{wang2011robust} and Elesin et al.\cite{elesin2012design}, and later in 3D by Men et al.\cite{men2014robust}. In the area of phononics, Sigmund and Jensen\cite{sigmund2003systematic} first used a gradient based topology optimization method to systematically design both phononic band-gap materials and structures. Gazonas et al.\cite{gazonas2006genetic} and Bilal and Hussein\cite{bilal2011ultrawide} implemented a genetic algorithm based topology optimization method for the design of phononic band gap structures. Alternate structure types and materials exhibiting  band-gap phenomenon have also been investigated. For  example, Jensen\cite{jensen2003phononic} considered mass-spring structures, Diaz et al.\cite{diaz2005design} designed band-gap grid structures, Halkj{\ae}r et al.\cite{halkjaer2006maximizing} maximized band gaps in plated structures, Olhoff et al.\cite{olhoff2012optimum} and Halkj{\ae}r and Sigmund\cite{halkjaer2004optimization} optimized band-gap beam structures.  Additionally, Vatanabe et al.\cite{vatanabe2014maximizing} maximized phononic band gaps in piezocomposite materials, Liu et al.\cite{liu2016systematic} explored the solid-solid phononic crystals for multiple separate band gaps with different polarizations, and Hedayatrasa et al. \cite{hedayatrasa2016optimal} optimized tunable phononic band gap plates under equibiaxial stretch.

Despite the considerable attention that topology optimization for 2-D phononic crystals has received, no work has been done on the topology optimization of 3-D phononic crystals. This is despite the potentially more useful nature of 3-D designs. Optimizing 2-D phononic crystals results in plate type designs which can have impacts on applications where wave propagation is constrained in 2-dimensions. However, wave propagation is inherently a 3-D phenomenon and optimization in 3-D can result in bulk materials which control wave propagation in all directions. This task is complicated by the challenging fact that phononic band structure evaluations are computationally expensive and that the computational complexity increases when band structure calculations are conducted repeatedly during the iterative process of optimization. The solution thus requires, first and foremost, an efficient phononic solver.

At this point there exist several numerical techniques for the evaluation of the phononic band structure. A good reference that discusses some of the most prominent techniques was published by Hussein\cite{hussein2009reduced} where the authors also presented a method of accelerating the existing algorithms through a secondary expansion. The Plane Wave Expansion \cite{kushwaha1994theory} method (PWE) and the Finite Element method \cite{hladky1991analysis,veres2012complexity} are two of the most commonly used solvers owing to the ease of their implementation and their  versatility. In this paper we have used a mixed variational method\cite{srivastava2014mixed,lu2016variational} to calculate phononic band structures. The mixed variational method is derived from the Hu-Washizu\cite{hu1955some,washizu1955variational} variational theorem and it admits variations on both the stress and displacement fields. The mixed method has been known to converge faster than Rayleigh quotient which forms the basis of the traditional displacement based Finite Element method\cite{babuvska1978numerical}. In a recently published comprehensive study we have shown that the mixed method also displays faster convergence than the PWE method \cite{lu2016variational}. In addition to using the mixed variational method as our solver we have achieved further computational accelerations by implementing it over distributed Graphical Processing Units \cite{srivastava2015gpu}. 

In this paper we have considered three main varieties of the cubic phononic crystal lattice (FCC, BCC, and SC). Our aim is to find  3-D topologies in a 2-material phononic crystal system that produce large all-angle, all-mode, normalized band gaps for each of the three symmetries considered. We evaluate the phononic band structures along the Irreducible Brillouin Zones (IBZ)\cite{brillouin2003wave,setyawan2010high} of the respective unit cells. The calculations are distributed over four compute nodes of a CPU-GPU hybrid cluster. We use a SIMP based topology optimization routine which is coupled with Heaviside projection for control over minimum feature sizes \cite{guest2004achieving}. The sensitivity analyses for the eigenvalue problem required for TO are also calculated in parallel and through a vectorization process. 
\section*{Results}
\subsection*{Phononic band structure calculation}\label{wave}
In the following calculations, the elastodynamic eigenvalue problem is formulated using the mixed variation method (Refer to Lu\cite{lu2016variational} for details). The propagation of waves in a three dimensional elastic medium is governed by 
\begin{eqnarray}
\label{eom}
\sigma_{mn,n}=-\lambda\varrho u_m,\\
\label{constitutiverelation}
u_{(j,k)}=D_{jkmn}\sigma_{mn},
\end{eqnarray}
where $\lambda=\omega^2$, $\boldsymbol{\sigma}$ and $\mathbf{u}$ are the space and time dependent stress tensor and displacement vector respectively, $\varrho$ is the mass density and $\mathbf{D}$ is the compliance tensor. The Latin indices vary from $1$ to $3$ and subject to the summation conventions unless otherwise indicated. By varying independently on the stress and displacement field and enforcing Bloch periodic boundary conditions, (\ref{eom}) and (\ref{constitutiverelation}) renders to the following functional stationary:
\begin{equation}\label{mixedvariation}
\lambda_{u\sigma}=\frac{\langle\sigma_{mn},u_{m,n}\rangle+\langle u_{j,k},\sigma_{jk}\rangle-\langle D_{jkmn}\sigma_{mn},\sigma_{jk}\rangle}{\langle\varrho u_m,u_m\rangle},
\end{equation}
and the minimum of the above quotient is the solutions to the phononic eigenvalue problem. This minimization problem can be solved by expanding the displacement and stress fields to satisfy the Bloch periodic boundary conditions. In this study, the trigonometric terms are used as test functions, $f^{\alpha\beta\gamma}(\mathbf{x})=\exp[\mathrm{i}2\pi K^{\alpha\beta\gamma}_k x_k ]$, where $K^{\alpha\beta\gamma}_k=T_{1k}(Q_1+\alpha)+T_{2k}(Q_2+\beta)+T_{3k}(Q_3+\gamma)$ and $Q_i$ are coordinates of the wave-vectors expressed in reciprocal lattice and $T$ transforms the orthogonal coordinate system to the primary lattice of the unit cell. By substituting test functions into the mixed variation formulation (\ref{mixedvariation}) and setting the derivative of $\lambda_{u\sigma}$ with respect to the unknown displacement and stress coefficients equal to zero and then eliminating the stress coefficients through matrix manipulation, we obtain the following general matrix form of the eigenvalue problem 
\begin{equation}\label{matrixform}
\mathbf{H}\boldsymbol{\Phi^{-1}}\mathbf{H}^*\mathbf{U}=\lambda_{u\sigma}\boldsymbol{\Omega}\mathbf{U},
\end{equation}
where elements of $\mathbf{U}$ are the field expansion coefficients and $(^*)$ indicates the complex conjugate operation of the matrix, and the expressions for the matrices are 
\begin{eqnarray}
\label{intH}
\mathbf{H}=\mathrm{i}2\pi K_k^{\theta\eta\xi}\int_V \exp\left[\mathrm{i}2\pi(K_l^{\alpha\beta\gamma}-K_l^{\theta\eta\xi})x_l\right]dV,\\
\label{intOmega}
\boldsymbol{\Omega}=\int_V \varrho(x_1,x_2,x_3)\exp\left[\mathrm{i}2\pi(K_l^{\alpha\beta\gamma}-K_l^{\theta\eta\xi})x_l\right]dV,\\
\label{intPhi}
\boldsymbol{\Phi}=\int_V D_{jkmn}(x_1,x_2,x_3)\exp\left[\mathrm{i}2\pi(K_l^{\alpha\beta\gamma}-K_l^{\theta\eta\xi})x_l\right]dV.
\end{eqnarray}
If $M$ trigonometric expansion terms are used, i.e. $\alpha$, $\beta$, $\gamma$ vary from $-M$ to $M$, then the size of the eigenvalue problem (\ref{matrixform}) will be $3(2M+1)^3\times3(2M+1)^3$ after considering the tensorial symmetries involved. Generally the mixed variation result will be neither upper nor lower bound of the eigenvalue solution\cite{nemat1975harmonic}. The above method has been proven to converge faster to the real solution, in terms of the matrix size and corresponding relative error, than typical band structure algorithms such as the FE method, the Rayleigh quotient, and the PWE method \cite{babuvska1978numerical,lu2016variational}. Detailed convergence studies were presented by Lu and Srivastava\cite{lu2016variational}. Comparison was made in terms of the number of basis terms (trigonometric or real) needed in the approximate expansion. In summary, it was found that the mixed-variational method results in a higher accuracy for a given size of the eigenvalue problem.
\subsection*{Computation complexity and efficiency}\label{hardware}
Phononic band structure computation accuracy directly influences the objective function evaluation and its efficiency determines the tractability of topology optimization implementation. When the number of trigonometric terms, $M$, increases in the mixed variational formulation, the band structure shows decreasing relative error\cite{lu2016variational}. Therefore, we prefer to use a large $M$ while keeping the matrix sizes manageable. In this study, 1029 trigonometric terms ($M=3$) are used to compute the band structure. The related matrices $\boldsymbol{\Omega}$, $\boldsymbol{\Phi}$ and $\mathbf{H}$ are of size $1029\times1029$, $2058\times2058$ and $1029\times2058$ respectively. Band structure evaluation has to be executed for 80 wave-vectors during each optimization iteration. The largest design domain in this study is a $48^3$ mesh, which gives rise to $110,592$ elements.  Gradient-based optimization requires sensitivity analysis, or specifically the derivatives of $\boldsymbol{\Omega}$ and $\boldsymbol{\Phi}$ with respect to the elemental design variables indicating material concentration.  This leads to 4.26TB float type data which must be manipulated during each iteration.
\begin{figure}[ht]
\centering
\includegraphics[scale=0.65]{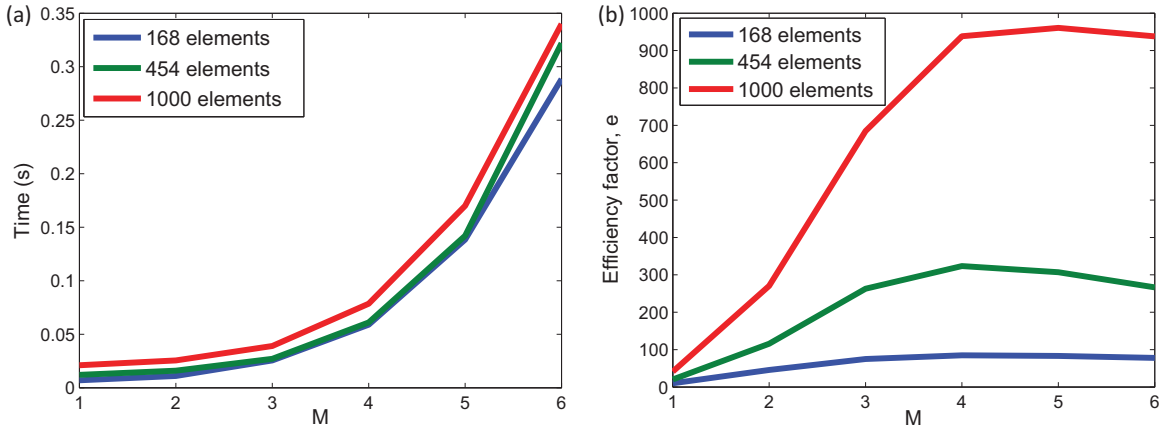}
\caption{(a) GPU accelerated computation time for eigenfrequencies solved at one wave-vector point. (b) Efficiency factor comparing the parallel formulation with the serial formulation.}
\label{efficiency}
\end{figure}
The computational cluster used for this work consists of 4 compute nodes, each of which has 4 NVIDIA GTX-780 graphic cards and 2 Intel(R) Xeon(R) E5-2630 v2 CPUs installed to form the mixed CPU-GPU architecture. Each GPU has 2304 CUDA cores and each CPU has 6 cores. In order to determine the parallel computation efficiency, we define an efficiency factor\cite{srivastava2015gpu} which measures the performance improvement through the parallel computations over serial computations in terms of the time it takes to do the same problem through the two methods:
\begin{equation}\label{efficiencyfactor}
e=\frac{t_{serial}}{t_{parallel}}
\end{equation}
Fig. \ref{efficiency}(a) shows the time taken for solving eigenfrequencies of a 2-D problem at one wave-vector and the efficiency factors are plotted in Fig. \ref{efficiency}(b). As  the number of trigonometric terms increases, the computation is 1000 times faster than the straightforward loop implementation. For a 2-D case, when M=3, the matrix size is $98 \times 98$ and the parallel computation time is about 0.05s when using 1000 elements. When using M=3 in a 3-D case, the computation will be 263 times more complex, due to the complexity of eigenvalue problem being $\mathit{O}(N^{2.37})$, where $N$ is the size of the matrix. Therefore, each eigenvalue problem takes about 13.2s to be solved and it takes about 65s to compute 80 eigenvalue problems in parallel on the 16GPUs to evaluate the necessary band structure.
\subsection*{A note on normalized band gaps}\label{note}
\begin{figure}[ht]
\centering
\includegraphics[scale=0.68]{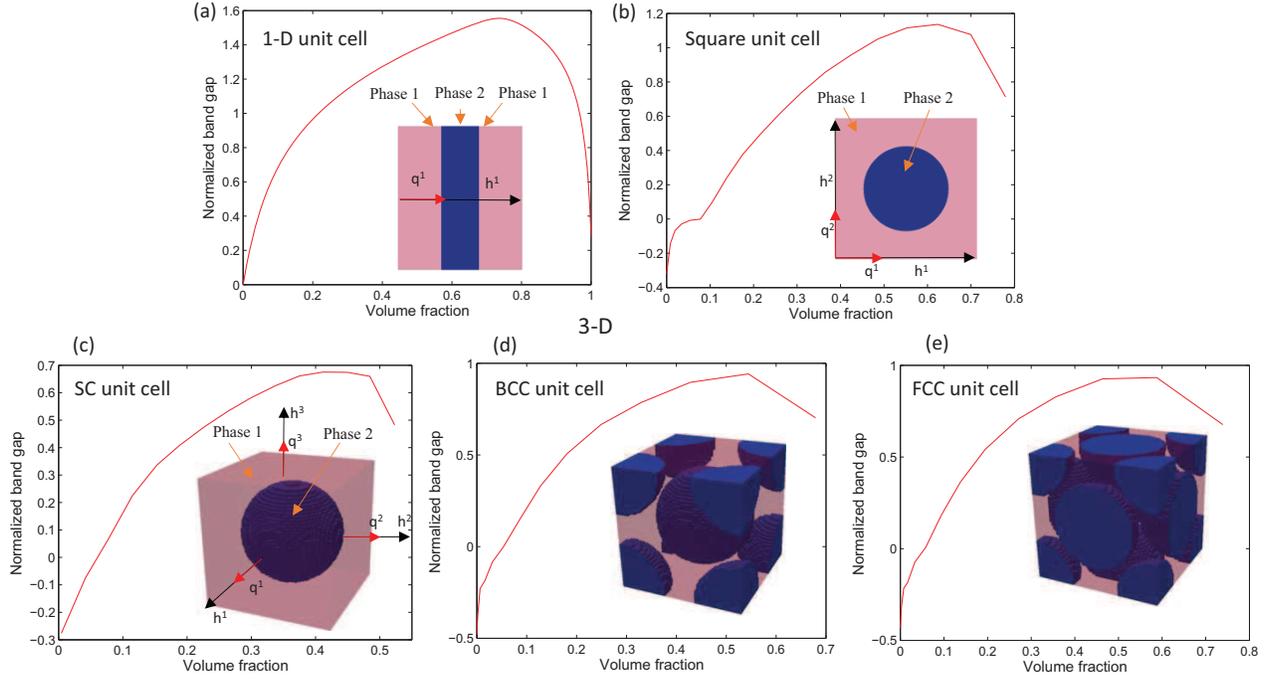}
\caption{The relation between the normalized band gap and volume fraction of the stiff material phase. (a) 1-D 2-phase layered composite; (b) 2-D 2-phase composite with circular inclusion; (c-e) 3-D 2-phase composite with spherical inclusions.}
\label{normalizedbandgap}
\end{figure}
Band structure is a plot of the phononic eigenfrequencies for wave-vectors which span the boundaries of IBZ. On this plot the normalized band gap is calculated by taking the ratio between the band gap width and the mid gap frequency. This metric is independent of unit cell scaling and is indicative of the tendency of a phononic crystal to stop all waves in all directions (for 3-D).
The simplest way to control normalized band gap sizes is to change the volume fractions of the material phases. Consider, for instance, simple phononic crystals made of tungsten carbide inclusions ($\varrho=13800kg/m^3$ , $E=387.5559GPa$, $\nu=0.3459$) in epoxy matrix ($\varrho=1180kg/m^3$, $E=4.3438GPa$, $\nu=0.3679$). 1-D phononic crystals are layered composites (Fig. \ref{normalizedbandgap}a). By changing the thickness of the material phases the maximal normalized band gap is seen to appear at a tungsten carbide volume fraction of 0.737, at which point the normalized band gap size is 155.5\%. Since thickness is the only design variable in this 1-D case this maximal gap size is also the global optimum for the material choices. Some optimization work has been reported for obtaining specific phononic bandstructure of interest, such as design of phononic filter\cite{hussein2007optimization} and solving for transient wave propagation problem\cite{dahl2008topology}. Circular inclusions inside a $1\times 1$ square unit cell are considered for a 2-D phononic crystal in Fig. \ref{normalizedbandgap}b. We observe that the maximum normalized band gap for this simple geometry has a value of 113.6\% and it occurs at a volume fraction of 0.624. Several studies have been conducted on exploring 2-D phononic topologies for larger bandgaps. For example, Bilal and Hussein\cite{bilal2011ultrawide} and Liu et al.\cite{liu2016systematic} have both achieved larger than 120\% normalized band gaps in 2-D. Figs. \ref{normalizedbandgap}(c-e) show analogous 3-D phononic cystals with $1\times1\times1$ cubic unit cells. The spherical inclusions lead to maximum normalized band gap values of 67.5\%, 94.2\% and 93.3\% at volume fraction 0.412, 0.545 and 0.588 for SC, BCC and FCC lattices respectively. We further note here that it becomes progressively more difficult to obtain large band gaps as we consider crystals of higher space dimensions.

The 1-D phononic crystal case is simple enough to admit theoretical arguments on largest possible bandgap values\cite{shmuel2016universality} and the 2-D case received research interest in terms of topology optimization studies. However, the 3-D phononic crystal band gap optimization has not yet been reported in literature. It is clear that formal topology optimization should be able to reveal crystals with larger stop bands than those produced by the simple inclusions in Fig. (\ref{normalizedbandgap}). 
\subsection*{Topology optimized structures}
For our study, topology optimization is performed over a $1cm^3$ cubic unit cell consisting of tungsten carbide and epoxy phases, allowing us to compare our results with previously published results\cite{page2005tunneling} of ultra-large band gap phononic crystals. The topological variable $\rho_e$ indicates the relative volume fraction of epoxy in each element, with $\rho_e=0$ indicating the element contains only tungsten carbide and $\rho_e=1$ indicating only epoxy.  During each iteration the stress and displacement fields are expanded using 1029 trigonometric terms ($M=3$). After the optimized solution is obtained a larger number of expansion terms ($M=4$) is used to calculate the final band structure at a higher accuracy. The objective function in this study is the normalized band gap between the 6th and 7th band, where a complete band gap opens naturally for a simple inclusion within a primitive cell. The longest single optimization iteration for the largest design domain, which consists of 48$^3$ elements without imposing symmetry, takes only 6.5min with $22.5\%$ of the time spent on band structure calculation and $45.5\%$ on sensitivity analysis. Design step direction calculation and data storage takes $32\%$ of the time in this case, however, if a smaller design domain is used, then this proportion will be smaller.

It should be noted that the considered topology optimization problem is nonconvex and thus the local minimum identified by the gradient-based optimizer is dependent on the topology used as the starting point for the optimization. Therefore, different material distributions are used as initial designs to help avoid getting trapped in low performance local minimum. Specifically, we have tested using spherical inclusions of various volume fractions and various homogeneous mixtures of tungsten carbide and epoxy as initial guesses.  Of course, as in most topology optimization problems, there is no guarantee that we have been able to identify a global optimum.
\subsubsection*{Simple cubic lattice}
\begin{figure}[ht]
\centering
\includegraphics[scale=0.65]{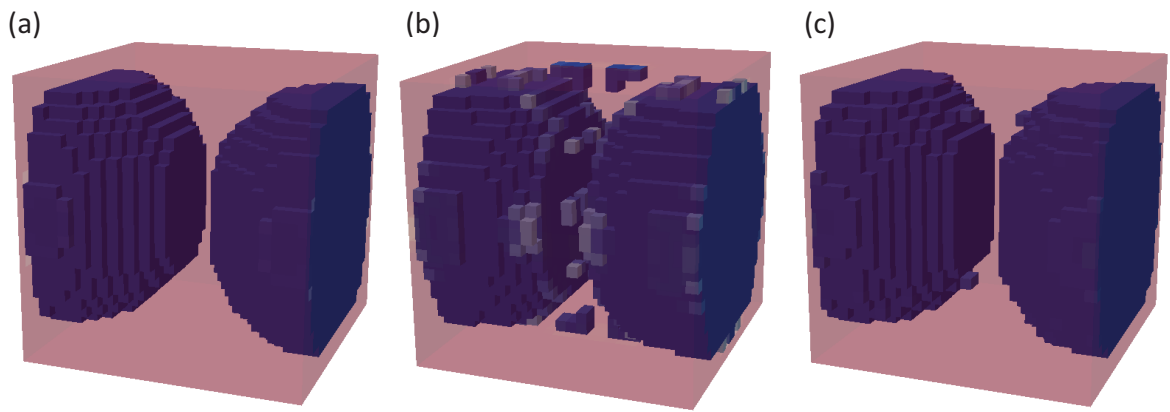}
\caption{Optimization results on $24^3$ mesh. Each case corresponds to different homogeneous initial designs having tungsten carbide percentages of (a)$35\%$, (b)$50\%$, (c)$65\%$.}
\label{sccoarse}
\end{figure}
As a first example, we consider the SC case defined on a relatively coarse mesh of $24^3$ elements with basic symmetries employed.  This means the optimization is  performed over 1/8 of the total number of elements with reflection symmetries assumed to generate the rest of the SC primitive cell. These initial results are presented in Fig. \ref{sccoarse} for different homogeneous initial distributions of material. When we take into account the periodicity of the unit cell, Figs. \ref{sccoarse}(a-c) show that for different homogeneous initial mixtures of tungsten carbide and epoxy, the optimization process converges to approximately the same topology after about 500 iterations. These solutions have similar normalized band gaps, which are $71.58\%$, $78.47\%$ and $70.47\%$ corresponding to tungsten carbide volume fractions 0.4374, 0.5015 and 0.4237, respectively.

\begin{figure}[ht]
\centering
\includegraphics[scale=.65]{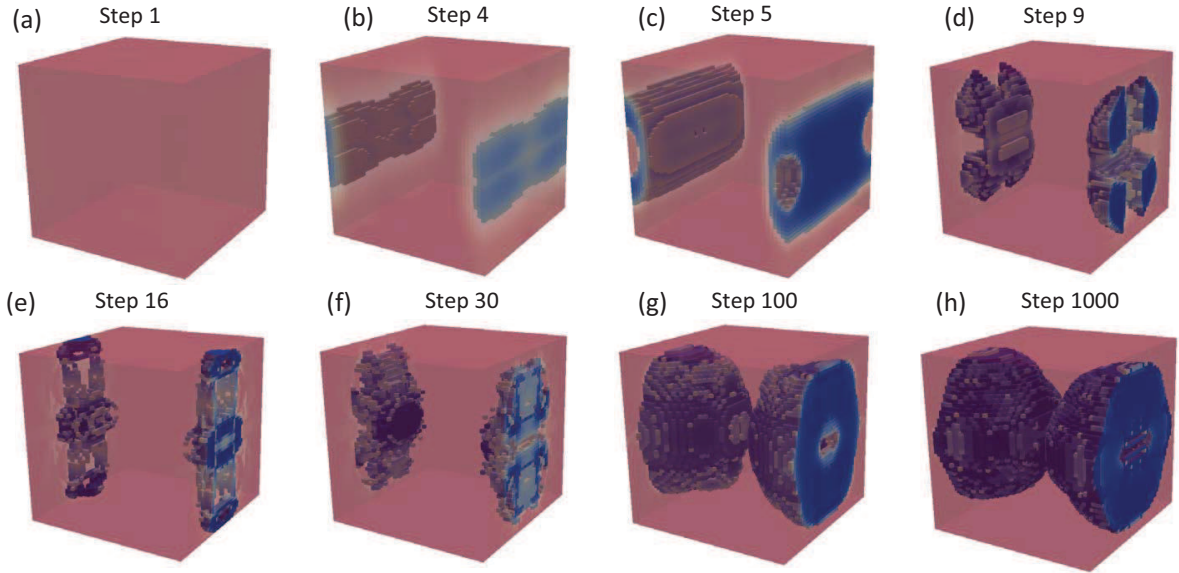}
\caption{Evolution of SC structure from homogeneous initial design on $48^3$ mesh.}
\label{scsteps}
\end{figure}

\begin{figure}[ht]
\centering
\includegraphics[scale=.65]{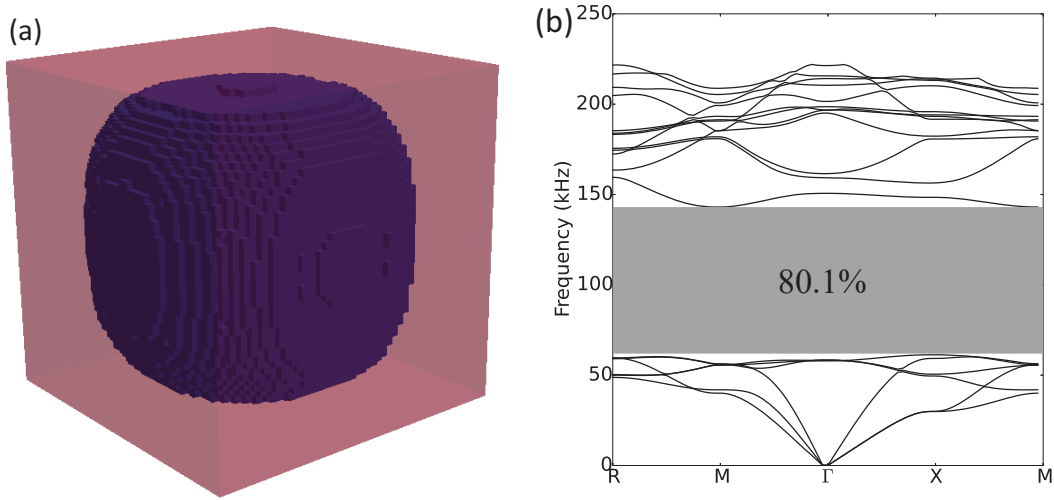}
\caption{(a) Optimized SC unit cell when using a single spherical inclusion as the initial design on 1/8 of the $48^3$ mesh due to imposed symmetries. (b) Corresponding band structure calculated using $M=4$.}
\label{optresultsSC}
\end{figure}

The SC problem using symmetry and a uniform initial distribution of  $\rho_e=0.5$ was re-solved on a finer mesh of $48^3$ elements. The optimization is performed over only 1/8 of the total number of elements due to imposed symmetries. The size of the data which is needed to be manipulated at each iteration is 545GB. The design evolution for this finer mesh case is shown in Fig. \ref{scsteps} and it is seen the algorithm evolves towards a single inclusion structure as in the previous coarse mesh cases.  The optimization converges after about 1000 iterations and the optimized inclusion structure is a cross between a sphere and a cube with some variation on the surface (Fig. \ref{optresultsSC}a). The structure has a normalized band gap ratio of 67.7\% and tungsten carbide volume fraction 0.3907.   

Since it is seen in both the coarse and fine mesh cases that the optimization process tends to evolve the topology into a structure that is equivalent to a single inclusion within a unit cell, we also used two variations of a centrally positioned stiff spherical inclusion as initial conditions for the optimization. The first case corresponds to a spherical inclusion with the largest volume fraction possible and the second corresponds to the one with largest normalized band gap (Fig. \ref{normalizedbandgap}c). The latter case resulted in a final structure with larger normalized band gap after convergence (about 1450 iterations). Fig. \ref{optresultsSC} show the results for the optimized SC design. The geometry of the inclusion, which has volume fraction 0.4771, resembles a cube with rounded corners. The corresponding stop band extends from $63.18kHz$ to $143.17kHz$ which is equivalent to a normalized band gap ratio of $80.1\%$. This is $18.67\%$ larger than the maximum possible SC band gap with a fully dense spherical inclusion.    

Finally, we note that we also attempted to relax the imposed symmetries and solve the SC case using an initial homogeneous distribution of $\rho_e=0.5$ in mesh of $48^3$ elements. This requires manipulation of 4.26TB float type data during each iteration and was thus significantly more computationally intensive than the previous case. The optimization tended towards highly asymmetric structures with non-existent or very small stop bands after several design iterations. This is likely due to the optimization process getting stuck in local minimas. The SC optimization problem without any additional symmetries is a very large computational problem with $>100,000$ variables. To adequately explore it we require faster computational algorithms and we expect that relaxing the symmetries will indeed result in larger bandgaps. However, current computational resources do not permit us to study this problem adequately.

\subsubsection*{Body-centered cubic lattice}
\begin{figure}[htp]
\centering
\includegraphics[scale=.8,clip=true]{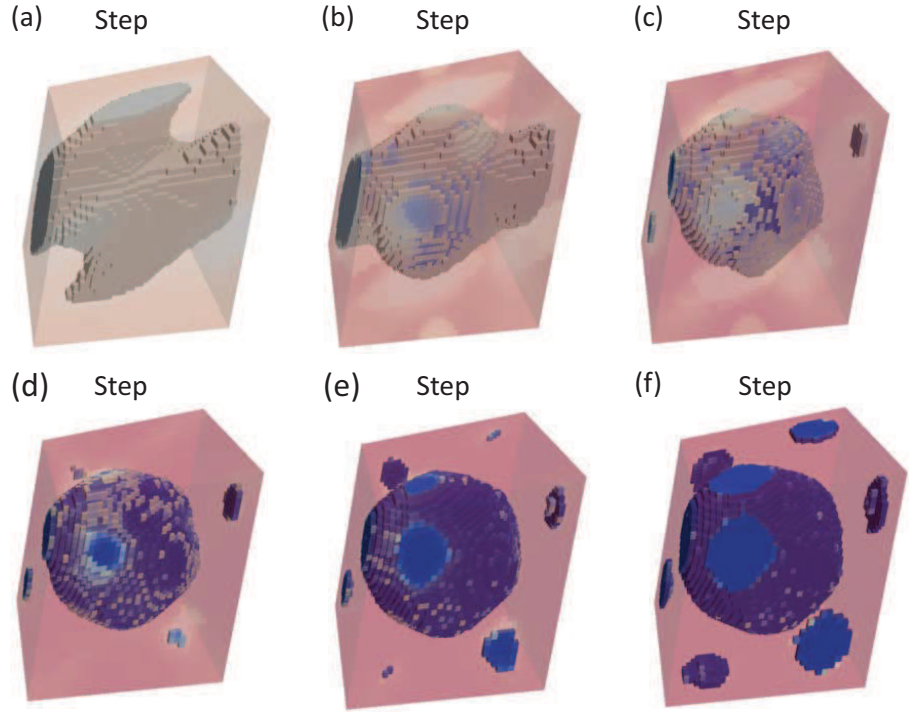}
\caption{Evolution of BCC structure from homogeneous initial design on $36^3$ mesh, where an equal fraction of tungsten carbide and epoxy is applied.}
\label{bccsteps}
\end{figure}
The BCC primitive cell is discretized into a $36^3$ mesh giving rise to a total of 46,656 elements. Optimization is performed without any additional symmetry constraints.  Fig. \ref{bccsteps} shows the design evolution, starting from a homogeneous initial design of $\rho_e=0.5$, progressing to a skewed cross-like structure, before converging after 186 iterations to a centrally-located large inclusion with smaller inclusions located near the unit cell corners as shown in Fig. \ref{bccsteps}(f). However, given the periodicity of the unit cell, the topology still corresponds to a single inclusion with the smaller inclusions joining with the large inclusion at appropriate locations in periodically repeated unit cells. The optimized structure features staggered inclusions  which have the general appearance of a sphere but with some variations on the surface.  The structure has normalized band gap ratio of 93.11\% with a tungsten carbide volume fraction 0.4417.
\begin{figure}[ht]
\centering
\includegraphics[scale=.65]{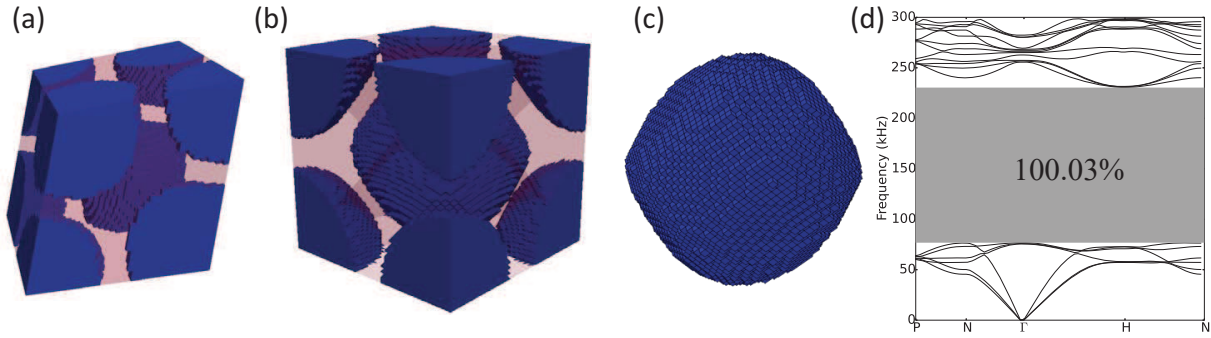}
\caption{(a) Optimized BCC primitive cell on $36^3$ mesh with portions of an inclusion located at the primitive cell corners. (b) Corresponding BCC unit cell extracted from the assembly of repeated BCC primitive cells. (c) Optimized tungsten carbide inclusion. (d) Corresponding band structure calculated using $M=4$.}
\label{optresultsBCC}
\end{figure}

We also considered two cases whose initial designs are spherical inclusions. As in the SC case, the first case corresponds to a spherical inclusion with the largest volume fraction possible and the second corresponds to the one with largest normalized band gap (Fig. \ref{normalizedbandgap}d). The former case resulted in a larger normalized band gap, converging after 287 iterations. Fig. \ref{optresultsBCC} shows the optimized BCC design. The optimized BCC primitive cell contains the complete geometric information of the single inclusion. This topology is then assembled to result in its corresponding unit cell by repeating the primitive cells based on the built-in translation symmetry of the BCC lattice (Fig. \ref{optresultsBCC}(b).) Much like the solution reported in Fig. \ref{bccsteps}(f), the resulting topology resembles a staggered pattern of tungsten carbide inclusions which have  the general appearance of spheres. However, large portions of the surfaces that face the body diagonals in these inclusions are flattened, as shown in Fig. \ref{optresultsBCC}(c). Although no internal symmetry constraint has been applied during the optimization process, the optimized inclusion structure has a centro-symmetric element arrangement. The optimized BCC structure has a band gap from $76.8kHz$ to $230.49kHz$, which leads to a normalized band gap value of $100.03\%$. The corresponding volume fraction of tungsten carbide in the optimized structure is 0.5475. This is in contrast with spherical inclusion results where the maximum possible normalized band gap is 94.2\% at a volume fraction of 0.5450.

\subsubsection*{Face-centered cubic lattice}
\begin{figure}[ht]
\centering
\includegraphics[scale=.68]{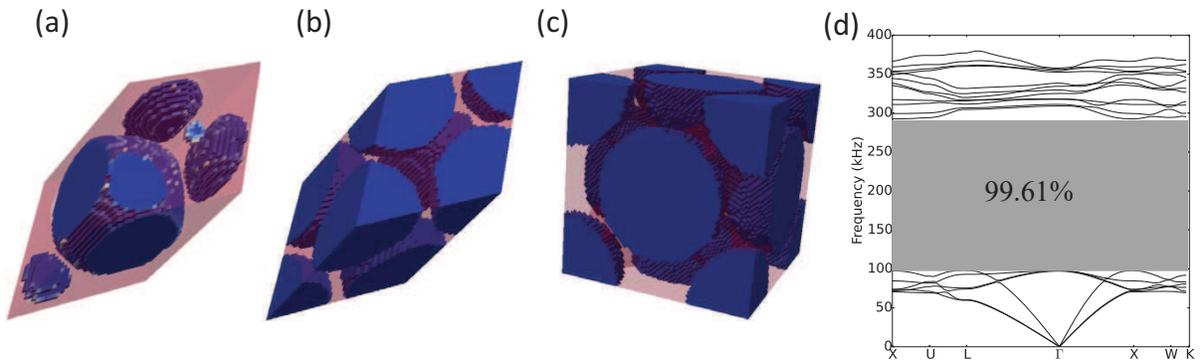}
\caption{(a) Optimized FCC primitive cell topology when using a homogeneous initial design on a $36^3$ mesh. (b) Optimized FCC primitive cell topology when using a single spherical inclusion as the initial guess. (c)  FCC unit cell assembled from the optimized FCC primitive cell in (b). (d) Corresponding band structure calculated using $M=4$.}
\label{optresultsFCC}
\end{figure} 
The FCC primitive cell is discretized into a $36^3$ mesh and optimization is performed without additional imposed symmetry constraints. The FCC case is solved using the same initial guesses as the BCC case: a homogeneous distribution with $\rho_e=0.5$ as well as a two spherical inclusion designs.  As in the BCC case, the topology optimized FCC solution found using a uniform initial distribution of material converges to a solution featuring a single inclusion shown in Fig. \ref{optresultsFCC}(a), with the general shape being close to a sphere. It has a normalized band gap of 95.24\% at a volume fraction 0.4861.  The solution found using an initial spherical distribution corresponding to the largest band gap in Fig. \ref{normalizedbandgap}(e) gave the best result, converging after 235 iterations to the optimized primitive and corresponding topologies shown in Figs. \ref{optresultsFCC}(b,c), respectively. Examining Fig. \ref{optresultsFCC}(c), it is clear that the shape of the tungsten carbide inclusion has the general appearance of a sphere with some variation on the surface. Although no internal symmetry constraint has been applied during the optimization process, it is seen that the optimized inclusion structure has a centro-symmetric element arrangement. The optimized FCC structure has a stop band from $98.03kHz$ to $292.55kHz$, which leads to a $99.61\%$ normalized band gap. The corresponding volume fraction of tungsten carbide in the optimized structure is 0.5502.

\begin{figure}[ht]
\centering
\includegraphics[scale=.7]{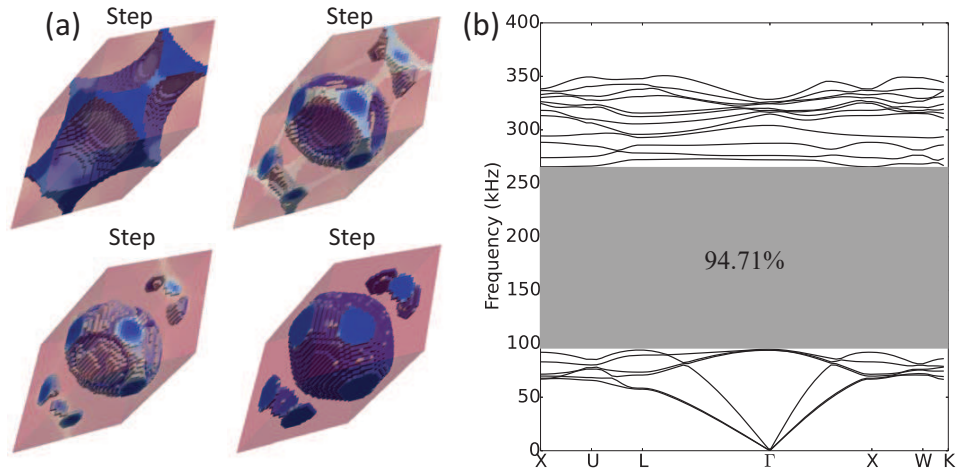}
\caption{(a) Evolution of the FCC lattice case with cermet topology as the initial design and  "network" topology as the converged result. (b) Band structure of the optimized network topology.}
\label{cermetsteps}
\end{figure}
According to Economou and Sigalas\cite{economou1993classical} the "cermet" topology, where inclusions consist of low-velocity materials surrounded by high-velocity matrix materials, is more favorable for the appearance of large elastic gaps. However, we could not find support for this in our studies. We found that beginning with cermet initial designs (Fig. \ref{cermetsteps}a), where an epoxy sphere is embedded in tungsten carbide matrix, the optimization process invariably steered towards network topologies in search of large band gaps. By step 6, the optimization progression has already converted the cermet initial design into a network design where a tungsten carbide inclusion is surrounded by epoxy matrix (Fig. \ref{cermetsteps}a). The optimization process converges after 134 iterations. The optimized structure has volume fraction 0.4752 and the stop band extends from $94.84kHz$ to $265.48kHz$, which leads to $94.71\%$ normalized band gap. Furthermore, in all our other optimization studies which involved beginning with a homogeneous initial design, the optimization process could have steered towards a cermet topology. However, it always resulted in network topologies.

\section*{Discussion}
In this paper we have presented the first ever topology optimization results for 3-D phononic crystals. Specifically our objective was to reveal 3-D phononic unit cells comprised of two material phases which display large all angle, all mode phononic band gaps. Specifically, optimized results for simple cubic, body centered cubic, and face centered cubic crystal structures made up of tungsten carbide and epoxy phases are presented. We have shown that for all these cases large phononic stop bands for bulk wave propagation can be achieved at lower than close packed configurations for spherical inclusions. A summary of the band gap results can be found in Table. \ref{results}, where we have compared our results with what is achievable through simple spherical inclusions. Specifically, it is possible to achieve normalized band gap of 67.5\%, 94.2\% and 93.3\% using tungsten carbide spheres in SC, BCC and FCC configurations respectively. Topology optimization shows that the SC result can be significantly improved to 80.1\% by modifying the shape of the inclusion, forming a cross between a sphere and a cube.  The BCC and FCC optimization results also improve over their spherical inclusion counterparts. Specifically, the BCC optimized structure shows a normalized bang gap greater than 100\%. Furthermore, it is interesting that the optimized structures shown in Fig. \ref{optresultsBCC}(a) and Fig. \ref{optresultsFCC}(b) achieve large complete band gaps at relatively small volume fractions of the stiff and heavy inclusion. For instance the BCC optimized structure achieves a large normalized band gap at a tungsten-carbide volume fraction of 0.5475. This is in comparison with a volume fraction of 0.74 which corresponds to the case where spheres are tightly packed in a FCC configuration. This results in a unit cell which is 23.1\% lighter than the close packed structure and still outperforms it in terms of its band gap size\cite{page2005tunneling}. In general, it is notable that largest normalized band gaps occur at less than close packed configuration both for spherical inclusions and optimized results.
\begin{table}[ht]
\centering
\caption{Summary of results.}
\label{results}
\begin{tabular}{lcccc}
\hline & \multicolumn{2}{c}{Spherical Inclusions} & \multicolumn{2}{c}{Optimized Inclusions}\\
\cline{2-3}  \cline{4-5}  
  & Normalized bang gap & Volume fraction & Normalized bang gap & Volume fraction \\
\hline
SC  & $67.5\%$ & 0.412 & $80.1\%$ & 0.4764 \\
BCC & $94.2\%$ & 0.545 & $100.0\%$ & 0.5475\\
FCC & $93.3\%$ & 0.588 & $99.6\%$ & 0.5502\\
\hline
\end{tabular}
\end{table}

3-D phononic band structure optimization is a highly computationally intensive task, enabled here through a GPU accelerated mixed variation method and mixed variation formulation based sensitivity analysis. For example, the 3-D SC lattice using a $48^3$ mesh translates into an optimization problem with over $100,000$ variables. Furthermore, with no assumed symmetry, the optimization process tends to get stuck in numerous local minima which correspond to highly asymmetric structures with small band gaps. For the SC case, therefore, we assume reflection symmetries over 1/8 of a unit cell. We studied various initial conditions with coarse ($24^3$) and fine ($48^3$) meshes. It was notable that almost all homogeneous initial conditions result in single contiguous inclusion topologies which are very similar to each other. All optimization runs which begin with spherical inclusions as their initial conditions also ended in simple contiguous inclusion topologies. This result was also noted for FCC and BCC runs. For both of these cases we used a $36^3$ mesh configured along the appropriate unit vectors. No further symmetries were assumed. In general, it was noted that homogeneous initial condition resulted in inclusions with more pronounced asymmetrical features. On the other hand, the cases with spherical inclusions as initial conditions resulted in more symmetrical inclusions with higher normalized band gap values. In our studies we could not find cases which support that cermet topology is more favorable in generating elastic band gaps. In fact, beginning with cermet initial designs, the optimization process invariably steered towards network topologies in search of large band gaps.

In summary, we have presented topology optimization results which reveal the largest as yet reported all angle all mode normalized bandgaps in 3-D phononic crystals. These results pertain to the two material combination of tungsten-carbide and epoxy. Other material combinations and/or optimizing band gaps between other bands (other than the 6th and 7th) will likely result in different optimized topologies and different normalized bandgaps.
\section*{Methods}
\subsection*{Topology optimization formulation}\label{Optimization}
In topology optimization of periodic composite materials, the goal is to optimize the distribution of two (or more) base material phases across the unit cell, which for finite element-based approaches, reduces to determining whether each element is to contain base material 1 or material 2 (see e.g., \cite{deaton2014survey,sigmund2013topology,cadman2013design,guest2016topology}).  This fundamentally is a binary (or integer) programming problem of extremely high dimension, motivating relaxation of the binary condition and representation of each element's material properties as a continuous combination of the two base materials. In order to obtain a binary design, penalization methods such as the Solid Isotropic Material with Penalization (SIMP) method are used to make mixtures of the two materials at a location inefficient.  Interestingly, Sigmund and Jensen\cite{sigmund2003systematic}  found  that the use of penalization is not required in the design of band-gap structures as sharp contrasts in stiffness (Young's modulus) are desirable to produce large band-gaps.  This allows use of a simple linear interpolation model for Young's modulus, given as
\begin{equation}\label{combinematerial}
E^{(e)}(\rho_e)=\rho_eE^1+\left(1-\rho_e\right)E^2,
\end{equation}
where $E^1$ and $E^2$ denote the Young’s modulus corresponding to the two base materials. We note this is equivalent to the SIMP interpolation for composites \cite{bendsoe1999material} with exponent penalty term set to one.  

The goal of the optimization is to maximize the gap between the $t$-th mode and $(t+1)$th mode, and thus we use the objective function given as
\begin{equation}\label{obj}
f(\rho_e)=2\frac{\min_k\lambda_{t+1}(\mathbf{k},\rho_e)-\max_k\lambda_t(\mathbf{k},\rho_e)}{\min_k\lambda_{t+1}(\mathbf{k},\rho_e)+\max_k\lambda_t(\mathbf{k},\rho_e)}
\end{equation}
and the resulting topology optimization formulation can be written as
\begin{equation}\label{topopt}
	\begin{aligned}
		\max\;\; & f(\rho_e)\\
		\mathrm{s.t.}\;\; & \left(\mathbf{H}\boldsymbol{\Phi}^{-1}\mathbf{H}^*-\lambda\boldsymbol{\Omega}\right)\mathbf{U}=0\\
		& 0\leq\rho_e\leq1,\; e=1,\cdots, N
	\end{aligned}
\end{equation}
Note that we do not impose a volume constraint and allow the algorithm to freely distribute the two base materials, although it would straightforward to restrict the design problem in that manner.

We want to emphasize that Eq. (\ref{topopt}) generally indicates an asymmetric eigenvalue problem. As a result the eigenvalues and corresponding eigenvectors are complex. Sensitivity calculations of complex eigenvalues require the use of normalized left and right eigenvectors, $\boldsymbol{\psi}$ and $\boldsymbol{\theta}$, in the sense that for each eigenvalue the corresponding pair of left and right eigenvectors should satisfy $\boldsymbol{\psi}^*\boldsymbol{\theta}=1$, such that
\begin{equation}\label{normalev}
\lambda=\boldsymbol{\psi}^*\left[\boldsymbol{\Omega}^{-1}\left(\mathbf{H}\boldsymbol{\Phi}^{-1}\mathbf{H}^*\right)\right]\boldsymbol{\theta}.
\end{equation}
Although the closed form expressions for $\boldsymbol{\Omega}^{-1}$ and $\boldsymbol{\Phi}^{-1}$ are implicit, based on the invertibility their derivatives with respect to design variable $\rho_e$ can be written as,
\begin{eqnarray}
\frac{\partial\boldsymbol{\Omega}^{-1}}{\partial\rho_{e}}=-\boldsymbol{\Omega}^{-1}\frac{\partial\boldsymbol{\Omega}}{\partial\rho_e}\boldsymbol{\Omega}^{-1}\\
\frac{\partial\boldsymbol{\Phi}^{-1}}{\partial\rho_{e}}=-\boldsymbol{\Phi}^{-1}\frac{\partial\boldsymbol{\Phi}}{\partial\rho_e}\boldsymbol{\Phi}^{-1}.
\end{eqnarray}
The sensitivity of the natural frequencies \cite{sigmund2003systematic} can be calculated by differentiating (\ref{normalev}) with respect to the design variables $\rho_e$, as follows
\begin{equation}\label{sensitivityew}
\frac{\partial\lambda}{\partial\rho_e}=\boldsymbol{\psi}^*\left[-\boldsymbol{\Omega}^{-1}\frac{\partial\boldsymbol{\Omega}}{\partial\rho_e}\boldsymbol{\Omega}^{-1}\left(\mathbf{H}\boldsymbol{\Phi}^{-1}\mathbf{H}^*\right)-\boldsymbol{\Omega}^{-1}\left(\mathbf{H}\boldsymbol{\Phi}^{-1}\frac{\partial\boldsymbol{\Phi}}{\partial\rho_e}\boldsymbol{\Phi}^{-1}\mathbf{H}^*\right)\right]\boldsymbol{\theta}.
\end{equation}
The sensitivity of the objective function can now be calculated by differentiating (\ref{obj}). 

Finally we note that the Heaviside Projection Method (HPM) \cite{guest2004achieving} is used within these formulations to control the minimum length scale of designed features. In particular, a reduced design variable field \cite{guest2010reducing} is adopted with design variables spaced at two times the finite element size. In HPM, the continuum design $\rho_e$ variables are expressed as a closed-form function of an independent design variable field without any other changes to the above equations.  For brevity, the details are omitted here; however, the reader is referred to Guest \cite{guest2009topology,guest2011eliminating} for full algorithmic details. Here, a continuation strategy is used on the Heaviside parameter, beginning at $\beta_{HPM}=1$. The projection radius is $1.4\%$ of the unit cell length. It should be noted that HPM is capable of controlling the minimum length scale of stiff and/or compliant designed features and thus preventing solution mesh dependency, although such dependencies have not been observed for phononic band-gap materials\cite{sigmund2003systematic}.
\subsection*{Vectorization and parallel computations}\label{GPU}
In order to calculate the entire band structure of a unit cell the matrices have to be assembled and the eigenvalues have to be calculated at multiple wave-vector points along the edge of the IBZ. This results in considerable computational complexity. However, since the assembly and eigenvalue solving processes are independent of each other they can be executed in parallel if the formulation is properly recast. The most basic computational unit in the formulation is the following integral in (\ref{intH}): 
\begin{equation}\label{loop}
I_{\alpha\beta\gamma\theta\eta\xi}=\int_V fdV=\int_V \exp\left[\mathrm{i}2\pi(K_l^{\alpha\beta\gamma}-K_l^{\theta\eta\xi})x_l\right]dV=\sum_{e=1}^{N}f^{(e)}(\alpha,\beta,\gamma,\theta,\eta,\xi)v^{(e)},
\end{equation} 
where $f^{(e)}$ is the evaluation of the integrand at the centroid of the $e$-th element, $v^{(e)}$ is the element volume and $N$ is the number of elements which discretizes the primitive cell. Recall that $K^{\alpha\beta\gamma}_l=T_{1l}(Q_1+\alpha)+T_{2l}(Q_2+\beta)+T_{3l}(Q_3+\gamma)$. The integrand can be expanded as
\begin{equation}
f^{(e)}(\alpha,\beta,\gamma,\theta,\eta,\xi)=\exp \left\lbrace \mathrm{i}2\pi\left[ T_{1l}(\alpha-\theta)+T_{2l}(\beta-\eta)+T_{3l}(\gamma-\xi)\right]x_l\right\rbrace.
\end{equation}
It is a constant matrix independent from the wave-vector coordinates $Q_i$. The above equation can be viewed as the outer product of $\mathbf{f}$ and its own complex  conjugate, where
\begin{equation}
\mathbf{f}=\exp \left[\mathrm{i}2\pi(T_{1l}\alpha+T_{2l}\beta+T_{3l}\gamma)x_l\right],
\end{equation}
is a vector of size $(2M+1)^3$. Now the integral (\ref{loop}) can be rewritten as
\begin{equation}\label{matrixofintegral}
I_{pq}=\sum_{e=1}^Nf^{(e)}_{pq}v^{(e)},
\end{equation}
where $p,q=1,2,\cdots,(2M+1)^3$. It is a constant global matrix which only needs to be calculated once. To compute these matrices using Graphical Processing Units we need to pass the vectors, volumes and centroids from the CPU to the GPU. On the GPU, the computation kernels are executed by a grid of thread blocks, where each thread has a unique id which corresponds to a set of indices $e$, $p$, $q$. Since the actual computation on each thread is relatively simple and many threads are operating in parallel, the method shows significantly reduced computation times compared to serial computations over a CPU. Furthermore, the band structure computations for different wave-vectors along the IBZ can be distributed over multiple GPUs in a distributed GPU cluster. In this case each compute node will solve only a part of the band structure thus decreasing the computation time further.

Sensitivity is calculated element-wise and therefore, parallel computation can significantly accelerate the process. Substituting (\ref{matrixofintegral}) into (\ref{intOmega}) and (\ref{intPhi}), the derivative of $\boldsymbol{\Omega}$ and $\boldsymbol{\Phi}$ with respect to the design variable of the $e$-th element, $\rho_{e}$, can be calculated by 
\begin{eqnarray}
\label{diffOmega}
\frac{\partial\boldsymbol{\Omega}}{\partial\rho_{e}}=\left(\varrho^1-\varrho^2\right)f^{(e)}_{pq}v^{(e)},\\
\label{diffPhi}
\frac{\partial\boldsymbol{\Phi}}{\partial\rho_{e}}=\left(D_{jkmn}^1-D_{jkmn}^2\right)f^{(e)}_{pq}v^{(e)},
\end{eqnarray}
where no summation is implied. We note that the sensitivity matrices of $\Omega$ and $\Phi$ of an element at location $x^{(e)}_l$ has the following transpose relation to the elements at the opposite location $-x^{(e)}_l$
\begin{eqnarray}
\frac{\partial\boldsymbol{\Omega}}{\partial\rho_{e}}\bigg|_{x_l=-x^{(e)}_l}=\left[\frac{\partial\boldsymbol{\Omega}}{\partial\rho_{e}}\bigg|_{x_l=x^{(e)}_l}\right]^T,\\
\frac{\partial\boldsymbol{\Phi}}{\partial\rho_{e}}\bigg|_{x_l=-x^{(e)}_l}=\left[\frac{\partial\boldsymbol{\Phi}}{\partial\rho_{e}}\bigg|_{x_l=x^{(e)}_l}\right]^T.
\end{eqnarray}
This is very helpful in reducing the computational complexity if the center of the design domain is positioned at the origin.
\subsection*{Study of cubic unit cells}\label{unitcell}
\begin{figure}[ht]
\centering
\includegraphics[scale=.65]{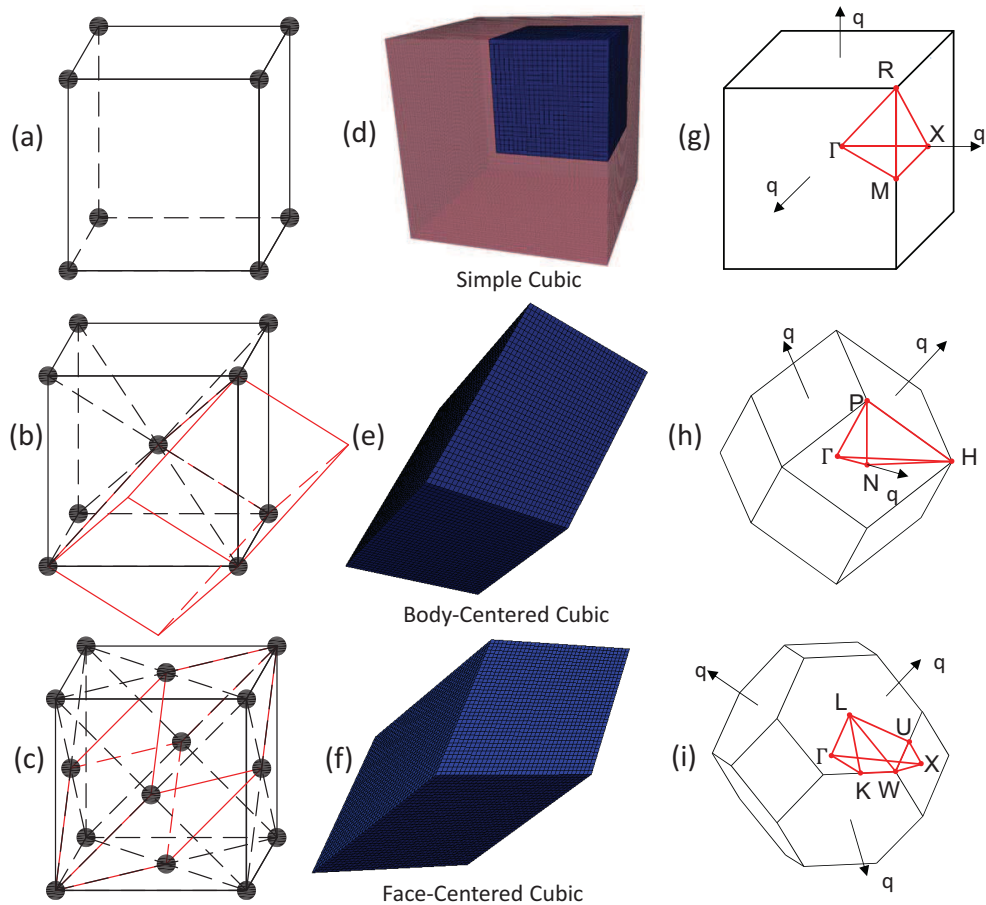}
\caption{(a), (b), (c) are the schematics of the SC, BCC, FCC lattices and the corresponding primitive cells. (d), (e), (f)  show the design domains that are actually used in the sensitivity analyses. (g), (h), (i) are the first Brillouin zones of the three cubic lattices. The red regions inside the Brillouin zones are the IBZs.}
\label{schematics}
\end{figure}

Cubic Bravais lattices are studied in this paper. The three varieties of cubic lattices considered here are the simple cubic (SC), the body-centered cubic (BCC), and the face-centered cubic (FCC) lattices \cite{hahn2005international}. The three lattices have their own variations based on different space groups.  For example, Maldovan et al. studied the photonic band gaps of 11 FCC structures\cite{maldovan2003exploring}. The discretization of the design domain should allow the formation of all possible variations during the optimization process, however, it is not very economic to use the entire cube as the design domain. In order to maintain the basic crystal structure during the optimization process, the geometry information has to be exactly the same on each lattice point. This can be realized by enforcing translation symmetries along their symmetry axes. As shown in Fig. \ref{schematics}(a-c), SC lattice's translation symmetry is along the lattice edges, whereas BCC and FCC lattice symmetry axes are along the body and face diagonals respectively. These translation symmetry axes are the vectors of the primitive cells which contains the geometry information of exactly one lattice point. Band structure calculations and sensitivity analyses are implemented on primitive cells. As demonstrated by Dong et al.\cite{dong2014topology,dong2015reducing}, reduction of symmetry is favorable in generating ultra-wide bandgaps, therefore, BCC and FCC primitive cells are discretized into $36^3$ elements and no further symmetries are assumed for them (Figs. \ref{schematics}e,f). The primitive cell of SC lattice is the cubic unit cell itself and it is discretized into $48^3$ elements. We had difficulty converging to a meaningful solution without imposing symmetry in the SC case where homogeneous material distribution is used as the initial design. Following Bilal and Hussein\cite{bilal2011ultrawide}, who implemented $C_{4v}$ rotational symmetry to their 2-D square lattice optimization, we  reduced the SC design domain to 1/8 of the primitive cell as shown in Fig. \ref{schematics}(d) and reconstruct the lattice by taking reflection symmetries, in which the reflecting mirrors are the orthogonal planes. Figs. \ref{schematics}(g-i) show the IBZ boundaries of SC, BCC, and FCC lattices respectively in the reciprocal space. During topology optimization the band structure is calculated for wave-vectors spanning these boundaries and then band gap location and size are extracted.

%\bibliography{ReferencesBib}

\begin{thebibliography}{10}
\expandafter\ifx\csname url\endcsname\relax
  \def\url#1{\texttt{#1}}\fi
\expandafter\ifx\csname urlprefix\endcsname\relax\def\urlprefix{URL }\fi
\expandafter\ifx\csname doiprefix\endcsname\relax\def\doiprefix{DOI: }\fi
\providecommand{\bibinfo}[2]{#2}
\providecommand{\eprint}[2][]{\url{#2}}

\bibitem{srivastava2015elastic}
\bibinfo{author}{Srivastava, A.}
\newblock \bibinfo{title}{Elastic metamaterials and dynamic homogenization: a
  review}.
\newblock \emph{\bibinfo{journal}{International Journal of Smart and Nano
  Materials}} \textbf{\bibinfo{volume}{6}}, \bibinfo{pages}{41--60}
  (\bibinfo{year}{2015}).

\bibitem{martinezsala1995sound}
\bibinfo{author}{Martinezsala, R.} \emph{et~al.}
\newblock \bibinfo{title}{Sound-attenuation by sculpture}.
\newblock \emph{\bibinfo{journal}{Nature}} \textbf{\bibinfo{volume}{378}},
  \bibinfo{pages}{241--241} (\bibinfo{year}{1995}).

\bibitem{bloch1928quantum}
\bibinfo{author}{Bloch, F.}
\newblock \bibinfo{title}{Quantum mechanics of electrons in crystal lattices}.
\newblock \emph{\bibinfo{journal}{Z. Phys}} \textbf{\bibinfo{volume}{52}},
  \bibinfo{pages}{555--600} (\bibinfo{year}{1928}).

\bibitem{ho1990existence}
\bibinfo{author}{Ho, K.~M.}, \bibinfo{author}{Chan, C.~T.} \&
  \bibinfo{author}{Soukoulis, C.~M.}
\newblock \bibinfo{title}{Existence of a photonic gap in periodic dielectric
  structures}.
\newblock \emph{\bibinfo{journal}{Physical Review Letters}}
  \textbf{\bibinfo{volume}{65}}, \bibinfo{pages}{3152--3155}
  (\bibinfo{year}{1990}).

\bibitem{khelif2003trapping}
\bibinfo{author}{Khelif, A.} \emph{et~al.}
\newblock \bibinfo{title}{Trapping and guiding of acoustic waves by defect
  modes in a full-band-gap ultrasonic crystal}.
\newblock \emph{\bibinfo{journal}{physical Review B}}
  \textbf{\bibinfo{volume}{68}}, \bibinfo{pages}{214301}
  (\bibinfo{year}{2003}).
\newblock \doiprefix 10.1103/PhysRevB.68.214301.

\bibitem{yang2002ultrasound}
\bibinfo{author}{Yang, S.} \emph{et~al.}
\newblock \bibinfo{title}{Ultrasound tunneling through 3d phononic crystals}.
\newblock \emph{\bibinfo{journal}{Physical review letters}}
  \textbf{\bibinfo{volume}{88}}, \bibinfo{pages}{104301}
  (\bibinfo{year}{2002}).
\newblock \doiprefix 10.1103/PhysRevLett.88.104301.

\bibitem{li2011tunable}
\bibinfo{author}{Li, X.-F.} \emph{et~al.}
\newblock \bibinfo{title}{Tunable unidirectional sound propagation through a
  sonic-crystal-based acoustic diode}.
\newblock \emph{\bibinfo{journal}{Physical review letters}}
  \textbf{\bibinfo{volume}{106}}, \bibinfo{pages}{084301}
  (\bibinfo{year}{2011}).
\newblock \doiprefix 10.1103/PhysRevLett.106.084301.

\bibitem{yang2004focusing}
\bibinfo{author}{Yang, S.} \emph{et~al.}
\newblock \bibinfo{title}{Focusing of sound in a 3d phononic crystal}.
\newblock \emph{\bibinfo{journal}{Physical review letters}}
  \textbf{\bibinfo{volume}{93}}, \bibinfo{pages}{024301}
  (\bibinfo{year}{2004}).
\newblock \doiprefix 10.1103/PhysRevLett.93.024301.

\bibitem{zen2014engineering}
\bibinfo{author}{Zen, N.}, \bibinfo{author}{Puurtinen, T.~A.},
  \bibinfo{author}{Isotalo, T.~J.}, \bibinfo{author}{Chaudhuri, S.} \&
  \bibinfo{author}{Maasilta, I.~J.}
\newblock \bibinfo{title}{Engineering thermal conductance using a
  two-dimensional phononic crystal}.
\newblock \emph{\bibinfo{journal}{Nature communications}}
  \textbf{\bibinfo{volume}{5}} (\bibinfo{year}{2014}).
\newblock \doiprefix 10.1038/ncomms4435.

\bibitem{nemat2015refraction}
\bibinfo{author}{Nemat-Nasser, S.}
\newblock \bibinfo{title}{Refraction characteristics of phononic crystals}.
\newblock \emph{\bibinfo{journal}{Acta Mechanica Sinica}}
  \textbf{\bibinfo{volume}{31}}, \bibinfo{pages}{481--493}
  (\bibinfo{year}{2015}).

\bibitem{nemat2015anti}
\bibinfo{author}{Nemat-Nasser, S.}
\newblock \bibinfo{title}{Anti-plane shear waves in periodic elastic
  composites: band structure and anomalous wave refraction}.
\newblock In \emph{\bibinfo{booktitle}{Proc. R. Soc. A}}, vol.
  \bibinfo{volume}{471}, \bibinfo{pages}{20150152} (\bibinfo{organization}{The
  Royal Society}, \bibinfo{year}{2015}).

\bibitem{srivastava2016metamaterial}
\bibinfo{author}{Srivastava, A.}
\newblock \bibinfo{title}{Metamaterial properties of periodic laminates}.
\newblock \emph{\bibinfo{journal}{Journal of the Mechanics and Physics of
  Solids}} \textbf{\bibinfo{volume}{96}}, \bibinfo{pages}{252--263}
  (\bibinfo{year}{2016}).

\bibitem{hussein2014dynamics}
\bibinfo{author}{Hussein, M.~I.}, \bibinfo{author}{Leamy, M.~J.} \&
  \bibinfo{author}{Ruzzene, M.}
\newblock \bibinfo{title}{Dynamics of phononic materials and structures:
  Historical origins, recent progress, and future outlook}.
\newblock \emph{\bibinfo{journal}{Applied Mechanics Reviews}}
  \textbf{\bibinfo{volume}{66}}, \bibinfo{pages}{040802}
  (\bibinfo{year}{2014}).
\newblock \doiprefix 10.1115/1.4026911.

\bibitem{deaton2014survey}
\bibinfo{author}{Deaton, J.~D.} \& \bibinfo{author}{Grandhi, R.~V.}
\newblock \bibinfo{title}{A survey of structural and multidisciplinary
  continuum topology optimization: post 2000}.
\newblock \emph{\bibinfo{journal}{Structural and Multidisciplinary
  Optimization}} \textbf{\bibinfo{volume}{49}}, \bibinfo{pages}{1--38}
  (\bibinfo{year}{2014}).

\bibitem{sigmund2013topology}
\bibinfo{author}{Sigmund, O.} \& \bibinfo{author}{Maute, K.}
\newblock \bibinfo{title}{Topology optimization approaches}.
\newblock \emph{\bibinfo{journal}{Structural and Multidisciplinary
  Optimization}} \textbf{\bibinfo{volume}{48}}, \bibinfo{pages}{1031--1055}
  (\bibinfo{year}{2013}).

\bibitem{cadman2013design}
\bibinfo{author}{Cadman, J.~E.}, \bibinfo{author}{Zhou, S.},
  \bibinfo{author}{Chen, Y.} \& \bibinfo{author}{Li, Q.}
\newblock \bibinfo{title}{On design of multi-functional microstructural
  materials}.
\newblock \emph{\bibinfo{journal}{Journal of Materials Science}}
  \textbf{\bibinfo{volume}{48}}, \bibinfo{pages}{51--66}
  (\bibinfo{year}{2013}).

\bibitem{guest2016topology}
\bibinfo{author}{Osanov, M.} \& \bibinfo{author}{Guest, J.}
\newblock \bibinfo{title}{Topology optimization for architected materials
  design}.
\newblock \emph{\bibinfo{journal}{Annual Review of Materials Research}}
  \textbf{\bibinfo{volume}{46}}, \bibinfo{pages}{211--233}
  (\bibinfo{year}{2016}).

\bibitem{bendsoe1989optimal}
\bibinfo{author}{Bends{\o}e, M.~P.}
\newblock \bibinfo{title}{Optimal shape design as a material distribution
  problem}.
\newblock \emph{\bibinfo{journal}{Structural Optimization}}
  \textbf{\bibinfo{volume}{1}}, \bibinfo{pages}{193--202}
  (\bibinfo{year}{1989}).

\bibitem{rozvany1991coc}
\bibinfo{author}{Rozvany, G.} \& \bibinfo{author}{Zhou, M.}
\newblock \bibinfo{title}{The coc algorithm, part i: cross-section optimization
  or sizing}.
\newblock \emph{\bibinfo{journal}{Computer Methods in Applied Mechanics and
  Engineering}} \textbf{\bibinfo{volume}{89}}, \bibinfo{pages}{281--308}
  (\bibinfo{year}{1991}).

\bibitem{asadpoure2015topology}
\bibinfo{author}{Asadpoure, A.} \& \bibinfo{author}{Valdevit, L.}
\newblock \bibinfo{title}{Topology optimization of lightweight periodic
  lattices under simultaneous compressive and shear stiffness constraints}.
\newblock \emph{\bibinfo{journal}{International Journal of Solids and
  Structures}} \textbf{\bibinfo{volume}{60}}, \bibinfo{pages}{1--16}
  (\bibinfo{year}{2015}).

\bibitem{challis2014high}
\bibinfo{author}{Challis, V.~J.}, \bibinfo{author}{Roberts, A.~P.} \&
  \bibinfo{author}{Grotowski, J.~F.}
\newblock \bibinfo{title}{High resolution topology optimization using graphics
  processing units (gpus)}.
\newblock \emph{\bibinfo{journal}{Structural and Multidisciplinary
  Optimization}} \textbf{\bibinfo{volume}{49}}, \bibinfo{pages}{315--325}
  (\bibinfo{year}{2014}).

\bibitem{dobson1999maximizing}
\bibinfo{author}{Dobson, D.~C.} \& \bibinfo{author}{Cox, S.~J.}
\newblock \bibinfo{title}{Maximizing band gaps in two-dimensional photonic
  crystals}.
\newblock \emph{\bibinfo{journal}{SIAM Journal on Applied Mathematics}}
  \textbf{\bibinfo{volume}{59}}, \bibinfo{pages}{2108--2120}
  (\bibinfo{year}{1999}).

\bibitem{cox2000band}
\bibinfo{author}{Cox, S.~J.} \& \bibinfo{author}{Dobson, D.~C.}
\newblock \bibinfo{title}{Band structure optimization of two-dimensional
  photonic crystals in h-polarization}.
\newblock \emph{\bibinfo{journal}{Journal of Computational Physics}}
  \textbf{\bibinfo{volume}{158}}, \bibinfo{pages}{214--224}
  (\bibinfo{year}{2000}).

\bibitem{jensen2004systematic}
\bibinfo{author}{Jensen, J.~S.} \& \bibinfo{author}{Sigmund, O.}
\newblock \bibinfo{title}{Systematic design of photonic crystal structures
  using topology optimization: Low-loss waveguide bends}.
\newblock \emph{\bibinfo{journal}{Applied Physics Letters}}
  \textbf{\bibinfo{volume}{84}}, \bibinfo{pages}{2022--2024}
  (\bibinfo{year}{2004}).

\bibitem{rupp2007design}
\bibinfo{author}{Rupp, C.~J.}, \bibinfo{author}{Evgrafov, A.},
  \bibinfo{author}{Maute, K.} \& \bibinfo{author}{Dunn, M.~L.}
\newblock \bibinfo{title}{Design of phononic materials/structures for surface
  wave devices using topology optimization}.
\newblock \emph{\bibinfo{journal}{Structural and Multidisciplinary
  Optimization}} \textbf{\bibinfo{volume}{34}}, \bibinfo{pages}{111--121}
  (\bibinfo{year}{2007}).

\bibitem{wang2011robust}
\bibinfo{author}{Wang, F.}, \bibinfo{author}{Jensen, J.~S.} \&
  \bibinfo{author}{Sigmund, O.}
\newblock \bibinfo{title}{Robust topology optimization of photonic crystal
  waveguides with tailored dispersion properties}.
\newblock \emph{\bibinfo{journal}{JOSA B}} \textbf{\bibinfo{volume}{28}},
  \bibinfo{pages}{387--397} (\bibinfo{year}{2011}).

\bibitem{elesin2012design}
\bibinfo{author}{Elesin, Y.}, \bibinfo{author}{Lazarov, B.~S.},
  \bibinfo{author}{Jensen, J.~S.} \& \bibinfo{author}{Sigmund, O.}
\newblock \bibinfo{title}{Design of robust and efficient photonic switches
  using topology optimization}.
\newblock \emph{\bibinfo{journal}{Photonics and nanostructures-Fundamentals and
  Applications}} \textbf{\bibinfo{volume}{10}}, \bibinfo{pages}{153--165}
  (\bibinfo{year}{2012}).

\bibitem{men2014robust}
\bibinfo{author}{Men, H.}, \bibinfo{author}{Lee, K.~Y.},
  \bibinfo{author}{Freund, R.~M.}, \bibinfo{author}{Peraire, J.} \&
  \bibinfo{author}{Johnson, S.~G.}
\newblock \bibinfo{title}{Robust topology optimization of three-dimensional
  photonic-crystal band-gap structures}.
\newblock \emph{\bibinfo{journal}{Optics express}}
  \textbf{\bibinfo{volume}{22}}, \bibinfo{pages}{22632--22648}
  (\bibinfo{year}{2014}).

\bibitem{sigmund2003systematic}
\bibinfo{author}{Sigmund, O.} \& \bibinfo{author}{Jensen, J.~S.}
\newblock \bibinfo{title}{Systematic design of phononic band--gap materials and
  structures by topology optimization}.
\newblock \emph{\bibinfo{journal}{Philosophical Transactions of the Royal
  Society of London. Series A: Mathematical, Physical and Engineering
  Sciences}} \textbf{\bibinfo{volume}{361}}, \bibinfo{pages}{1001--1019}
  (\bibinfo{year}{2003}).

\bibitem{gazonas2006genetic}
\bibinfo{author}{Gazonas, G.~A.}, \bibinfo{author}{Weile, D.~S.},
  \bibinfo{author}{Wildman, R.} \& \bibinfo{author}{Mohan, A.}
\newblock \bibinfo{title}{Genetic algorithm optimization of phononic bandgap
  structures}.
\newblock \emph{\bibinfo{journal}{International journal of solids and
  structures}} \textbf{\bibinfo{volume}{43}}, \bibinfo{pages}{5851--5866}
  (\bibinfo{year}{2006}).

\bibitem{bilal2011ultrawide}
\bibinfo{author}{Bilal, O.~R.} \& \bibinfo{author}{Hussein, M.~I.}
\newblock \bibinfo{title}{Ultrawide phononic band gap for combined in-plane and
  out-of-plane waves}.
\newblock \emph{\bibinfo{journal}{Physical Review E}}
  \textbf{\bibinfo{volume}{84}}, \bibinfo{pages}{065701}
  (\bibinfo{year}{2011}).
\newblock \doiprefix 10.1103/PhysRevE.84.065701.

\bibitem{jensen2003phononic}
\bibinfo{author}{Jensen, J.~S.}
\newblock \bibinfo{title}{Phononic band gaps and vibrations in one-and
  two-dimensional mass--spring structures}.
\newblock \emph{\bibinfo{journal}{Journal of Sound and Vibration}}
  \textbf{\bibinfo{volume}{266}}, \bibinfo{pages}{1053--1078}
  (\bibinfo{year}{2003}).

\bibitem{diaz2005design}
\bibinfo{author}{Diaz, A.}, \bibinfo{author}{Haddow, A.} \&
  \bibinfo{author}{Ma, L.}
\newblock \bibinfo{title}{Design of band-gap grid structures}.
\newblock \emph{\bibinfo{journal}{Structural and Multidisciplinary
  Optimization}} \textbf{\bibinfo{volume}{29}}, \bibinfo{pages}{418--431}
  (\bibinfo{year}{2005}).

\bibitem{halkjaer2006maximizing}
\bibinfo{author}{Halkj{\ae}r, S.}, \bibinfo{author}{Sigmund, O.} \&
  \bibinfo{author}{Jensen, J.~S.}
\newblock \bibinfo{title}{Maximizing band gaps in plate structures}.
\newblock \emph{\bibinfo{journal}{Structural and Multidisciplinary
  Optimization}} \textbf{\bibinfo{volume}{32}}, \bibinfo{pages}{263--275}
  (\bibinfo{year}{2006}).

\bibitem{olhoff2012optimum}
\bibinfo{author}{Olhoff, N.}, \bibinfo{author}{Niu, B.} \&
  \bibinfo{author}{Cheng, G.}
\newblock \bibinfo{title}{Optimum design of band-gap beam structures}.
\newblock \emph{\bibinfo{journal}{International Journal of Solids and
  Structures}} \textbf{\bibinfo{volume}{49}}, \bibinfo{pages}{3158--3169}
  (\bibinfo{year}{2012}).

\bibitem{halkjaer2004optimization}
\bibinfo{author}{Halkj{\ae}r, S.} \& \bibinfo{author}{Sigmund, O.}
\newblock \bibinfo{title}{Optimization of beam properties with respect to
  maximum band-gap}.
\newblock In \emph{\bibinfo{booktitle}{Mechanics of the 21st Century,
  Procedings of 21st International Congress of Theoretical and Applied
  Mechanics}} (\bibinfo{organization}{IUTAM, Warsaw, Poland},
  \bibinfo{year}{2004}).

\bibitem{vatanabe2014maximizing}
\bibinfo{author}{Vatanabe, S.~L.}, \bibinfo{author}{Paulino, G.~H.} \&
  \bibinfo{author}{Silva, E.~C.}
\newblock \bibinfo{title}{Maximizing phononic band gaps in piezocomposite
  materials by means of topology optimization}.
\newblock \emph{\bibinfo{journal}{The Journal of the Acoustical Society of
  America}} \textbf{\bibinfo{volume}{136}}, \bibinfo{pages}{494--501}
  (\bibinfo{year}{2014}).

\bibitem{liu2016systematic}
\bibinfo{author}{Liu, Z.-F.}, \bibinfo{author}{Wu, B.} \& \bibinfo{author}{He,
  C.-F.}
\newblock \bibinfo{title}{Systematic topology optimization of solid--solid
  phononic crystals for multiple separate band-gaps with different
  polarizations}.
\newblock \emph{\bibinfo{journal}{Ultrasonics}} \textbf{\bibinfo{volume}{65}},
  \bibinfo{pages}{249--257} (\bibinfo{year}{2016}).

\bibitem{hedayatrasa2016optimal}
\bibinfo{author}{Hedayatrasa, S.}, \bibinfo{author}{Abhary, K.},
  \bibinfo{author}{Uddin, M.} \& \bibinfo{author}{Guest, J.~K.}
\newblock \bibinfo{title}{Optimal design of tunable phononic bandgap plates
  under equibiaxial stretch}.
\newblock \emph{\bibinfo{journal}{Smart Materials and Structures}}
  \textbf{\bibinfo{volume}{25}}, \bibinfo{pages}{055025}
  (\bibinfo{year}{2016}).
\newblock \doiprefix 10.1088/0964-1726/25/5/055025.

\bibitem{hussein2009reduced}
\bibinfo{author}{Hussein, M.}
\newblock \bibinfo{title}{Reduced bloch mode expansion for periodic media band
  structure calculations}.
\newblock \emph{\bibinfo{journal}{Proceedings of the Royal Society A:
  Mathematical, Physical and Engineering Science}}
  \textbf{\bibinfo{volume}{465}}, \bibinfo{pages}{2825--2848}
  (\bibinfo{year}{2009}).

\bibitem{kushwaha1994theory}
\bibinfo{author}{Kushwaha, M.}, \bibinfo{author}{Halevi, P.},
  \bibinfo{author}{Martinez, G.}, \bibinfo{author}{Dobrzynski, L.} \&
  \bibinfo{author}{Djafari-Rouhani, B.}
\newblock \bibinfo{title}{Theory of acoustic band structure of periodic elastic
  composites}.
\newblock \emph{\bibinfo{journal}{Physical Review B}}
  \textbf{\bibinfo{volume}{49}}, \bibinfo{pages}{2313--2322}
  (\bibinfo{year}{1994}).

\bibitem{hladky1991analysis}
\bibinfo{author}{Hladky-Hennion, A.-C.} \& \bibinfo{author}{Decarpigny, J.-N.}
\newblock \bibinfo{title}{Analysis of the scattering of a plane acoustic wave
  by a doubly periodic structure using the finite element method: Application
  to alberich anechoic coatings}.
\newblock \emph{\bibinfo{journal}{The Journal of the Acoustical Society of
  America}} \textbf{\bibinfo{volume}{90}}, \bibinfo{pages}{3356--3367}
  (\bibinfo{year}{1991}).

\bibitem{veres2012complexity}
\bibinfo{author}{Veres, I.~A.} \& \bibinfo{author}{Berer, T.}
\newblock \bibinfo{title}{Complexity of band structures: Semi-analytical finite
  element analysis of one-dimensional surface phononic crystals}.
\newblock \emph{\bibinfo{journal}{Physical Review B}}
  \textbf{\bibinfo{volume}{86}}, \bibinfo{pages}{104304}
  (\bibinfo{year}{2012}).
\newblock \doiprefix 10.1103/PhysRevB.86.104304.

\bibitem{srivastava2014mixed}
\bibinfo{author}{Srivastava, A.} \& \bibinfo{author}{Nemat-Nasser, S.}
\newblock \bibinfo{title}{Mixed-variational formulation for phononic
  band-structure calculation of arbitrary unit cells}.
\newblock \emph{\bibinfo{journal}{Mechanics of Materials}}
  \textbf{\bibinfo{volume}{74}}, \bibinfo{pages}{67--75}
  (\bibinfo{year}{2014}).

\bibitem{lu2016variational}
\bibinfo{author}{Lu, Y.} \& \bibinfo{author}{Srivastava, A.}
\newblock \bibinfo{title}{Variational methods for phononic calculations}.
\newblock \emph{\bibinfo{journal}{Wave Motion}} \textbf{\bibinfo{volume}{60}},
  \bibinfo{pages}{46--61} (\bibinfo{year}{2016}).

\bibitem{hu1955some}
\bibinfo{author}{Hu, H.-C.}
\newblock \bibinfo{title}{On some variational principles in the theory of
  elasticity and the theory of plasticity}.
\newblock \emph{\bibinfo{journal}{Scientia Sinica}}
  \textbf{\bibinfo{volume}{4}}, \bibinfo{pages}{33--54} (\bibinfo{year}{1955}).

\bibitem{washizu1955variational}
\bibinfo{author}{Washizu, K.}
\newblock \bibinfo{title}{On the variational principles of elasticity and
  plasticity}.
\newblock \bibinfo{type}{Tech. Rep.} \bibinfo{number}{25-18},
  \bibinfo{institution}{Aeroelastic and Structures Research Laboratory},
  \bibinfo{address}{MIT Press, Cambridge} (\bibinfo{year}{1955}).

\bibitem{babuvska1978numerical}
\bibinfo{author}{Babu{\v{s}}ka, I.} \& \bibinfo{author}{Osborn, J.}
\newblock \bibinfo{title}{Numerical treatment of eigenvalue problems for
  differential equations with discontinuous coefficients}.
\newblock \emph{\bibinfo{journal}{Mathematics of Computation}}
  \textbf{\bibinfo{volume}{32}}, \bibinfo{pages}{991--1023}
  (\bibinfo{year}{1978}).

\bibitem{srivastava2015gpu}
\bibinfo{author}{Srivastava, A.}
\newblock \bibinfo{title}{Gpu accelerated variational methods for fast phononic
  eigenvalue solutions}.
\newblock In \emph{\bibinfo{booktitle}{SPIE Smart Structures and Materials+
  Nondestructive Evaluation and Health Monitoring}}, \bibinfo{pages}{94381F}
  (\bibinfo{organization}{International Society for Optics and Photonics},
  \bibinfo{year}{2015}).

\bibitem{brillouin2003wave}
\bibinfo{author}{Brillouin, L.}
\newblock \emph{\bibinfo{title}{Wave propagation in periodic structures:
  electric filters and crystal lattices}} (\bibinfo{publisher}{Courier
  Corporation}, \bibinfo{year}{2003}).

\bibitem{setyawan2010high}
\bibinfo{author}{Setyawan, W.} \& \bibinfo{author}{Curtarolo, S.}
\newblock \bibinfo{title}{High-throughput electronic band structure
  calculations: Challenges and tools}.
\newblock \emph{\bibinfo{journal}{Computational Materials Science}}
  \textbf{\bibinfo{volume}{49}}, \bibinfo{pages}{299--312}
  (\bibinfo{year}{2010}).

\bibitem{guest2004achieving}
\bibinfo{author}{Guest, J.~K.}, \bibinfo{author}{Pr{\'e}vost, J.~H.} \&
  \bibinfo{author}{Belytschko, T.}
\newblock \bibinfo{title}{Achieving minimum length scale in topology
  optimization using nodal design variables and projection functions}.
\newblock \emph{\bibinfo{journal}{International journal for numerical methods
  in engineering}} \textbf{\bibinfo{volume}{61}}, \bibinfo{pages}{238--254}
  (\bibinfo{year}{2004}).

\bibitem{nemat1975harmonic}
\bibinfo{author}{Nemat-Nasser, S.}, \bibinfo{author}{Fu, F.} \&
  \bibinfo{author}{Minagawa, S.}
\newblock \bibinfo{title}{Harmonic waves in one-, two-and three-dimensional
  composites: Bounds for eigenfrequencies}.
\newblock \emph{\bibinfo{journal}{International Journal of Solids and
  Structures}} \textbf{\bibinfo{volume}{11}}, \bibinfo{pages}{617--642}
  (\bibinfo{year}{1975}).

\bibitem{hussein2007optimization}
\bibinfo{author}{Hussein, M.~I.} \& \bibinfo{author}{El-Beltagy, M.~A.}
\newblock \bibinfo{title}{Optimization of phononic filters via genetic
  algorithms}.
\newblock In \emph{\bibinfo{booktitle}{Journal of Physics: Conference Series}},
  vol.~\bibinfo{volume}{92}, \bibinfo{pages}{012110}
  (\bibinfo{organization}{IOP Publishing}, \bibinfo{year}{2007}).

\bibitem{dahl2008topology}
\bibinfo{author}{Dahl, J.}, \bibinfo{author}{Jensen, J.~S.} \&
  \bibinfo{author}{Sigmund, O.}
\newblock \bibinfo{title}{Topology optimization for transient wave propagation
  problems in one dimension}.
\newblock \emph{\bibinfo{journal}{Structural and Multidisciplinary
  Optimization}} \textbf{\bibinfo{volume}{36}}, \bibinfo{pages}{585--595}
  (\bibinfo{year}{2008}).

\bibitem{shmuel2016universality}
\bibinfo{author}{Shmuel, G.} \& \bibinfo{author}{Band, R.}
\newblock \bibinfo{title}{Universality of the frequency spectrum of laminates}.
\newblock \emph{\bibinfo{journal}{Journal of the Mechanics and Physics of
  Solids}} \textbf{\bibinfo{volume}{92}}, \bibinfo{pages}{127--136}
  (\bibinfo{year}{2016}).

\bibitem{page2005tunneling}
\bibinfo{author}{Page, J.~H.} \emph{et~al.}
\newblock \bibinfo{title}{Tunneling and dispersion in 3d phononic crystals}.
\newblock \emph{\bibinfo{journal}{Zeitschrift f{\"u}r
  Kristallographie-Crystalline Materials}} \textbf{\bibinfo{volume}{220}},
  \bibinfo{pages}{859--870} (\bibinfo{year}{2005}).

\bibitem{economou1993classical}
\bibinfo{author}{Economou, E.} \& \bibinfo{author}{Sigalas, M.}
\newblock \bibinfo{title}{Classical wave propagation in periodic structures:
  Cermet versus network topology}.
\newblock \emph{\bibinfo{journal}{Physical Review B}}
  \textbf{\bibinfo{volume}{48}}, \bibinfo{pages}{13434--13438}
  (\bibinfo{year}{1993}).

\bibitem{bendsoe1999material}
\bibinfo{author}{Bends{\o}e, M.~P.} \& \bibinfo{author}{Sigmund, O.}
\newblock \bibinfo{title}{Material interpolation schemes in topology
  optimization}.
\newblock \emph{\bibinfo{journal}{Archive of applied mechanics}}
  \textbf{\bibinfo{volume}{69}}, \bibinfo{pages}{635--654}
  (\bibinfo{year}{1999}).

\bibitem{guest2010reducing}
\bibinfo{author}{Guest, J.~K.} \& \bibinfo{author}{Smith~Genut, L.~C.}
\newblock \bibinfo{title}{Reducing dimensionality in topology optimization
  using adaptive design variable fields}.
\newblock \emph{\bibinfo{journal}{International Journal for Numerical Methods
  in Engineering}} \textbf{\bibinfo{volume}{81}}, \bibinfo{pages}{1019--1045}
  (\bibinfo{year}{2010}).

\bibitem{guest2009topology}
\bibinfo{author}{Guest, J.~K.}
\newblock \bibinfo{title}{Topology optimization with multiple phase
  projection}.
\newblock \emph{\bibinfo{journal}{Computer Methods in Applied Mechanics and
  Engineering}} \textbf{\bibinfo{volume}{199}}, \bibinfo{pages}{123--135}
  (\bibinfo{year}{2009}).

\bibitem{guest2011eliminating}
\bibinfo{author}{Guest, J.~K.}, \bibinfo{author}{Asadpoure, A.} \&
  \bibinfo{author}{Ha, S.-H.}
\newblock \bibinfo{title}{Eliminating beta-continuation from heaviside
  projection and density filter algorithms}.
\newblock \emph{\bibinfo{journal}{Structural and Multidisciplinary
  Optimization}} \textbf{\bibinfo{volume}{44}}, \bibinfo{pages}{443--453}
  (\bibinfo{year}{2011}).

\bibitem{hahn2005international}
\bibinfo{author}{Hahn, T.}
\newblock \emph{\bibinfo{title}{International Tables for Crystallography,
  Space-Group Symmetry}} (\bibinfo{publisher}{Springer Science \& Business
  Media}, \bibinfo{year}{2005}).

\bibitem{maldovan2003exploring}
\bibinfo{author}{Maldovan, M.}, \bibinfo{author}{Ullal, C.~K.},
  \bibinfo{author}{Carter, W.~C.} \& \bibinfo{author}{Thomas, E.~L.}
\newblock \bibinfo{title}{Exploring for 3d photonic bandgap structures in the
  11 fcc space groups}.
\newblock \emph{\bibinfo{journal}{Nature materials}}
  \textbf{\bibinfo{volume}{2}}, \bibinfo{pages}{664--667}
  (\bibinfo{year}{2003}).

\bibitem{dong2014topology}
\bibinfo{author}{Dong, H.-W.}, \bibinfo{author}{Su, X.-X.},
  \bibinfo{author}{Wang, Y.-S.} \& \bibinfo{author}{Zhang, C.}
\newblock \bibinfo{title}{Topology optimization of two-dimensional asymmetrical
  phononic crystals}.
\newblock \emph{\bibinfo{journal}{Physics Letters A}}
  \textbf{\bibinfo{volume}{378}}, \bibinfo{pages}{434--441}
  (\bibinfo{year}{2014}).

\bibitem{dong2015reducing}
\bibinfo{author}{Dong, H.-W.}, \bibinfo{author}{Wang, Y.-S.},
  \bibinfo{author}{Wang, Y.-F.} \& \bibinfo{author}{Zhang, C.}
\newblock \bibinfo{title}{Reducing symmetry in topology optimization of
  two-dimensional porous phononic crystals}.
\newblock \emph{\bibinfo{journal}{AIP Advances}} \textbf{\bibinfo{volume}{5}},
  \bibinfo{pages}{117149} (\bibinfo{year}{2015}).
\newblock \doiprefix 10.1063/1.4936640.

\end{thebibliography}

\section*{Acknowledgements}

A.S. would like to acknowledge the support of UCSD/ONR W91CRB-10-1-0006 award to the Illinois Institute of Technology and NSF CAREER grant \#
1554033 to the Illinois Institute of Technology. J.K.G. would like to acknowledge the support of NSF Grants 1400394 and 1538367 to Johns Hopkins University.

\section*{Author contributions statement}

Y.L. and Y.Y. conducted the computational simulations and optimization, Y.L analyzed the results. J.K.G. and A.S. supervised the study. All authors contributed to writing the manuscript. 

\section*{Additional information}

\textbf{Competing financial interests:} The authors declare no competing financial interests. 

\end{document}